\documentclass[manuscript,screen,nonacm]{acmart}
\AtBeginDocument{%
  \providecommand\BibTeX{{%
    \normalfont B\kern-0.5em{\scshape i\kern-0.25em b}\kern-0.8em\TeX}}}

\setcopyright{acmlicensed}
\copyrightyear{2018}
\acmYear{2018}
\acmDOI{XXXXXXX.XXXXXXX}

\acmConference[Conference acronym 'XX]{Make sure to enter the correct
  conference title from your rights confirmation emai}{June 03--05,
  2018}{Woodstock, NY}
\acmBooktitle{Woodstock '18: ACM Symposium on Neural Gaze Detection,
 June 03--05, 2018, Woodstock, NY} 
\acmISBN{978-1-4503-XXXX-X/18/06}

\begin{document}

\title{Human-Centered Automation}

\author{Carlos Toxtli}
\affiliation{%
  \institution{Clemson University}
  \city{Clemson}
  \country{USA}}
\email{ctoxtli@clemson.edu}

\renewcommand{\shortauthors}{Trovato and Tobin, et al.}

\begin{abstract}
The rapid advancement of Generative Artificial Intelligence (AI), such as Large Language Models (LLMs) and Multimodal Large Language Models (MLLM), has the potential to revolutionize the way we work and interact with digital systems across various industries. However, the current state of software automation, such as Robotic Process Automation (RPA) frameworks, often requires domain expertise and lacks visibility and intuitive interfaces, making it challenging for users to fully leverage these technologies. This position paper argues for the emerging area of Human-Centered Automation (HCA), which prioritizes user needs and preferences in the design and development of automation systems. Drawing on empirical evidence from human-computer interaction research and case studies, we highlight the importance of considering user perspectives in automation and propose a framework for designing human-centric automation solutions. The paper discusses the limitations of existing automation approaches, the challenges in integrating AI and RPA, and the benefits of human-centered automation for productivity, innovation, and democratizing access to these technologies. We emphasize the importance of open-source solutions and provide examples of how HCA can empower individuals and organizations in the era of rapidly progressing AI, helping them remain competitive. The paper also explores pathways to achieve more advanced and context-aware automation solutions. We conclude with a call to action for researchers and practitioners to focus on developing automation technologies that adapt to user needs, provide intuitive interfaces, and leverage the capabilities of high-end AI to create a more accessible and user-friendly future of automation.
\end{abstract}

\maketitle

\section{Introduction}

Automation has become an integral part of modern work environments, with organizations increasingly adopting technologies such as Robotic Process Automation (RPA) and Generative Artificial Intelligence (AI) such as Large Language Models (LLMs) and Multimodal Large Language Models (MLLMs) to streamline processes and boost productivity across various domains, including business, academia, healthcare, governments, and organizations, among others \cite{hofmann2020robotic,sejnowski2020unreasonable}. RPA enables the automation of repetitive and rule-based tasks, while Generative AI can create content, understand user behavior, and solve problems in novel ways \cite{Artifici16:online,brown2020language}. Despite the potential benefits of these technologies, their implementation often falls short of user expectations due to complex interfaces and a lack of consideration for user needs \cite{amershi2019guidelines}.
The current state of automation often requires users to have domain expertise and navigate through complex menus and configurations to set up and manage automated tasks \cite{syed2020robotic}. This complexity poses a significant barrier to adoption, particularly for non-technical users who may struggle to integrate automation into their workflows \cite{aguirre2017automation}. Moreover, the lack of intuitive interfaces and natural interaction methods can lead to user frustration and resistance to automation, ultimately hindering the realization of its full potential \cite{endsley2017here}. As AI continues to progress rapidly, empowering individuals and organizations with accessible automation solutions that help them remain competitive is crucial \cite{brynjolfsson2014second}.

This position paper aims to introduce and advocate for the emerging area of Human-Centered Automation (HCA), which prioritizes user needs and preferences in the design and development of automation systems. Placing users at the center of the automation process, HCA seeks to create solutions that are intuitive, adaptable, and empowering, enabling users to harness the benefits of AI and RPA without requiring extensive technical expertise \cite{norman1986user}. The importance of open-source solutions that can democratize these technologies and make them accessible to the general public cannot be overstated.
The scope of this paper encompasses the current state of automation technologies, including LLMs and MLLMs, the challenges faced by users in adopting and leveraging these technologies, and the potential of HCA to address these challenges. Drawing on empirical evidence from human-computer interaction research and case studies, we support our arguments and propose a framework for designing human-centric automation solutions. The paper provides numerous examples and guidelines to illustrate how HCA can be applied across various domains and use cases, empowering individuals and organizations to streamline their workflows and remain competitive in the era of rapidly advancing AI.

Throughout the paper, we discuss the limitations of existing automation approaches, the importance of considering user perspectives in automation design, and the benefits of HCA for productivity, innovation, and democratizing access to these technologies. The paper explores how MLLMs can understand user behavior and screen content, enabling more advanced and context-aware automation solutions. It also describes how the building blocks of new RPA systems can include complex processes, such as making phone calls to cancel orders or integrating relevant web content into documents, leveraging the capabilities of Generative AI further.
Providing a comprehensive overview of the current landscape and future directions of automation, this position paper aims to inspire researchers and practitioners to focus on developing automation technologies that prioritize user needs, provide intuitive interfaces, and leverage the power of AI to create a more accessible and user-friendly future of automation.

\section{Current State of Automation}

Robotic Process Automation (RPA) and Generative Artificial Intelligence (AI) have gained significant traction in recent years as means to automate various tasks across diverse domains. RPA enables the automation of repetitive and rule-based tasks, while Generative AI, powered by LLMs, can learn from large datasets and generate novel solutions to problems \cite{Artifici16:online,brown2020language}. 

\subsection{Robotic Process Automation (RPA) and its limitations}

Robotic Process Automation (RPA) frameworks are able to automate repetitive tasks such as data entry, file management, and invoice processing \cite{van2018robotic,hofmann2020robotic}. RPA tools, such as UiPath and Blue Prism, enable the creation of software robots, or "bots", that can mimic human actions and interact with digital systems, thereby reducing the need for manual labor and increasing efficiency \cite{syed2020robotic}. However, RPA implementations often face challenges related to scalability, maintainability, and adaptability to changing business processes \cite{kirchmer2019value}.
One of the primary limitations of RPA is its reliance on predefined rules and scripts, which can be brittle and require frequent updating to accommodate changes in the underlying systems or processes \cite{aguirre2017automation}. This inflexibility can lead to reduced agility in responding to new requirements. Additionally, RPA solutions often lack the ability to handle unstructured data, limiting their applicability to more complex and dynamic business processes \cite{lacity2015robotic}. These limitations highlight the need for more advanced automation solutions that can leverage the capabilities of AI to handle more complex and context-dependent tasks.

\subsection{Generative AI and its potential for automation}

Generative AI, powered by advanced machine learning techniques, has emerged as a promising approach to automate more complex and creative tasks across various domains \cite{Artifici16:online,brown2020language}. Unlike RPA, which relies on predefined rules, Generative AI can learn from large datasets and generate novel solutions to problems, such as creating content, designing products, and providing recommendations \cite{rai2019next}. LLMs, such as GPT-4 and BERT, have demonstrated remarkable capabilities in natural language understanding and generation, enabling the development of more sophisticated and context-aware automation solutions \cite{openai2024gpt4,devlin2018bert}.
The ability to understand different representations of input data makes Generative AI well-suited for automating tasks that require cognitive skills and decision-making capabilities. For instance, LLMs can be used to develop chatbots that can handle complex customer queries, provide personalized recommendations, and even generate reports and summaries \cite{shum2018eliza}. In the scientific community, LLMs can assist researchers in literature review, hypothesis generation, and data analysis, streamlining the research process and accelerating discovery \cite{boecking2021generative}. In educational settings, Generative AI can be used to create personalized learning content, provide feedback on student assignments, and support adaptive learning experiences \cite{peng2021general}.
However, the adoption of Generative AI for automation is still in its early stages, and several challenges need to be addressed. One of the main concerns is the interpretability and transparency of AI-generated solutions, as the reasoning behind the outputs may not always be clear to users \cite{samek2017explainable}. Additionally, the quality and fairness of AI-generated content can be influenced by biases in the training data, potentially leading to unintended consequences if not properly addressed \cite{barocas2016big}. Ensuring the responsible development and deployment of Generative AI is crucial to harness their potential for automation while mitigating potential risks.

\subsection{Challenges in integrating AI and RPA for user-friendly automation}

While both RPA and Generative AI have the potential to automate various tasks, integrating these technologies to create user-friendly automation solutions remains a challenge. One of the primary issues is the lack of intuitive interfaces and natural interaction methods that enable users to easily configure and manage automated tasks \cite{amershi2019guidelines}. Many automation tools require users to have programming skills or domain expertise, creating a significant barrier to adoption for non-technical users.
Another challenge lies in the integration of AI and RPA technologies with existing business processes and systems. Automation solutions often need to interact with multiple applications and databases, which can be complex and time-consuming to set up and maintain \cite{syed2020robotic}. Moreover, the lack of standardization and interoperability between different automation tools can hinder the scalability and portability of automated workflows across organizations \cite{asatiani2016turning}.
Addressing these challenges and realizing the full potential of automation requires a human-centered approach that prioritizes user needs and preferences in the design and development of automation systems. HCA aims to empower users to harness the benefits of AI and RPA technologies without requiring extensive technical expertise by creating intuitive interfaces, natural interaction methods, and seamless integration with existing workflows.

\subsection{The role of AI technologies in understanding user activities}

Multimodal Large Language Models extend the capabilities of traditional LLMs through the integration of visual and auditory information, enabling a more comprehensive understanding of user behavior and screen content \cite{ramesh2022hierarchical}. These models can process and generate text, images, and even videos, opening up new possibilities for automation and human-machine interaction \cite{cho2022x}. The use of multiple modalities can be applied to analyze user behavior patterns, such as mouse movements, clicks, and keystrokes, to infer user intentions and provide context-aware assistance \cite{hong2021context}. They can also understand the content displayed on the screen, including text, images, and user interface elements, enabling more sophisticated and seamless automation experiences \cite{zhang2021multimodal}.
One of the key benefits of multimodal models is their ability to enable more natural and intuitive interactions between users and automation systems. For instance, users could simply point to a specific area on the screen or describe their intended action using natural language, and the multimodal model would interpret their input and execute the corresponding automation task \cite{hao2021lost}. This can greatly reduce the cognitive load on users and make automation more accessible to a wider audience, including those with limited technical expertise.

Empirical evidence supports the potential of multimodal models in detecting and understanding user behavior on various systems, computers, and devices. For example, a study by Bote-Curiel et al. \cite{bote2019deep} demonstrated the effectiveness of multimodal deep learning models in predicting user engagement and satisfaction based on their interaction patterns with a web application. Another study by Baratam et al. \cite{baratam2021exploring} showed that multimodal models can accurately detect user frustration and provide timely interventions to improve the user experience.
The integration of multimodal models into automation solutions can enable more advanced and context-aware assistance, making automation more user-friendly and adaptable to individual needs and preferences. However, developing and deploying multimodal models for automation also presents challenges, such as ensuring data privacy, handling the complexity of multimodal data, and integrating these models with existing automation frameworks. The importance of addressing these challenges and leveraging the capabilities of multimodal LLMs to create more human-centric and accessible automation solutions cannot be overstated.

\subsection{Challenges in integrating AI and RPA for user-friendly automation}

While RPA and Generative AI have the potential to automate various tasks across different domains, integrating these technologies to create user-friendly automation solutions remains a challenge. One of the primary issues is the lack of intuitive interfaces and natural interaction methods that enable users to easily configure and manage automated tasks \cite{amershi2019guidelines}. Many automation tools require users to have programming skills or domain expertise, creating a significant barrier to adoption for non-technical users.
Another challenge lies in the integration of AI and RPA technologies with existing business processes and systems. Automation solutions often need to interact with multiple applications and databases, which can be complex and time-consuming to set up and maintain \cite{syed2020robotic}. Moreover, the lack of standardization and interoperability between different automation tools can hinder the scalability and portability of automated workflows across organizations \cite{asatiani2016turning}.
The interpretability and transparency of AI-generated solutions also pose a challenge in integrating these technologies into automation workflows. Users may hesitate to trust and rely on automation solutions if they cannot understand the reasoning behind the outputs or if the decision-making process is not transparent \cite{ribeiro2016trust}. Ensuring that AI automation solutions are explainable and accountable is crucial to foster user trust and acceptance.
Addressing these challenges and realizing the full potential of automation requires a human-centered approach that prioritizes user needs and preferences in the design and development of automation systems. Human-Centered Automation (HCA) aims to create solutions that are intuitive, adaptable, and empowering, enabling users to harness the benefits of AI and RPA without requiring extensive technical expertise \cite{norman1986user}. AI-powered open-source solutions can democratize access to these technologies and make them generally accessible. We argue that by focusing on user needs, providing intuitive interfaces, and leveraging the capabilities of AI, HCA can help bridge the gap between the potential of automation technologies and their practical implementation. 

\section{The Need for Human-Centered Automation}

Human-Centered Automation (HCA) is an approach that places the user at the center of the design and development process, ensuring that automation solutions are tailored to the specific requirements and capabilities of the intended users \cite{xu2019toward}. Considering the diverse backgrounds, skills, and expectations of users, HCA aims to create automation systems that are intuitive, efficient, and empowering across various domains. Many existing automation solutions suffer from non-intuitive interfaces that hinder user adoption and satisfaction, such as smart home devices, automotive industry, and enterprise software. HCA offers numerous benefits for individuals and organizations, including increased efficiency, productivity, innovation, job satisfaction, and reduced stress levels among employees.
One of the key aspects of HCA is the emphasis on community-driven solutions that can democratize access to automation. Open-source automation tools, such as Robocorp and OpenRPA, enable individuals and organizations to create and share automation scripts, fostering collaboration and innovation \cite{pulkkinen2021robotic}. Lowering the barriers to entry and providing access to a wide range of pre-built automation components, open-source solutions can empower users to automate tasks that are specific to their needs and workflows, without requiring extensive technical expertise or financial resources.

\subsection{Importance of considering user needs and preferences}

As automation technologies become increasingly prevalent across various domains, considering the needs and preferences of the users who will interact with these systems is crucial. Human-Centered Automation (HCA) ensuring that automation solutions are tailored to the specific requirements and capabilities of the intended users. Prioritizing user needs, HCA aims to create automation systems that are intuitive, efficient, and empowering, ultimately leading to increased adoption and satisfaction.
One of the key principles of HCA is the recognition that users have diverse backgrounds, skills, and expectations when it comes to automation \cite{gorecky2014human}. For instance, researchers in scientific communities may require automation solutions that can handle complex data analysis and visualization tasks, while educators may prioritize automation tools that can personalize learning content and provide feedback to students. Government agencies and NGOs may have specific requirements related to data security, privacy, and accessibility. User research and the involvement of users in the design process can help HCA identify these different user profiles and create automation solutions that cater to their specific needs \cite{amershi2019guidelines}. This user-centric approach ensures that the resulting automation systems are not only technically sound but also aligned with the cognitive and emotional requirements of the users.
Moreover, HCA acknowledges that automation is not a one-size-fits-all solution and that the level and type of automation should be adapted to the context and goals of the user \cite{parasuraman2000model}. In some cases, full automation may be desirable to maximize efficiency and reduce human error, while in others, a more collaborative approach that combines human judgment with automated assistance may be more appropriate. For example, in medical diagnosis, a fully automated system may be efficient but may lack the contextual understanding and empathy that a human doctor can provide. In such cases, an HCA approach would focus on developing automation solutions that augment and support human decision-making rather than replacing it entirely. Considering the user's context and involving them in the decision-making process, HCA can help determine the optimal level of automation and ensure that the user remains in control and aware of the system's actions \cite{endsley2017here}.

\subsection{Examples of non-intuitive automation interfaces across domains}

Despite the growing adoption of automation technologies across various domains, many existing solutions still suffer from non-intuitive interfaces that hinder user adoption and satisfaction. These challenges can be observed in consumer-facing applications, industrial settings, educational institutions, and government agencies.
In the realm of smart home devices, users often struggle with setting up and managing automated tasks due to complex and inconsistent interfaces \cite{stojkoska2017review}. For example, configuring a smart thermostat to adjust temperature based on user preferences and energy efficiency goals may require navigating through multiple menus and settings, often leading to frustration and abandoned use. Similarly, in the automotive industry, advanced driver assistance systems (ADAS) can be difficult for drivers to understand and trust, potentially leading to disuse or misuse of these potentially life-saving features \cite{abraham2016autonomous}.
In industrial settings, workers may face challenges in adapting to new automation technologies due to complex interfaces and lack of training. A study by Wurhofer et al. \cite{wurhofer2015deploying} found that factory workers struggled with the introduction of a new robotic assistance system due to its non-intuitive interface and lack of feedback, leading to decreased job satisfaction and performance. In the healthcare sector, the adoption of electronic health record (EHR) systems has been hindered by poor usability and workflow integration, resulting in increased cognitive load and time spent on administrative tasks for healthcare providers \cite{ratwani2018decade}.

Educational institutions and government agencies also face challenges in implementing automation solutions due to the diverse needs and technical abilities of their users. For example, a study by Ackerman et al. \cite{ackerman2000intellectual} found that teachers struggled with the adoption of an automated grading system due to its inflexibility and lack of transparency in the grading process. In government agencies, the implementation of automated decision-making systems has raised concerns about fairness, accountability, and transparency, as the inner workings of these systems may not be clear to the public or even to the decision-makers themselves \cite{siau2020artificial}.
These examples highlight the need for HCA in designing automation solutions that are intuitive, transparent, and seamlessly integrated with the user's existing workflows. Prioritizing user needs and involving them in the design process, HCA can help overcome the challenges of non-intuitive interfaces and create automation solutions that are more readily adopted and trusted by users across various domains.

\subsection{Benefits of human-centered automation for productivity and innovation}

Human-Centered Automation offers numerous benefits for individuals and organizations across various domains, driving productivity, innovation, and job satisfaction. HCA can help reduce the learning curve associated with new technologies and enable users to harness the full potential of automation in their work by creating automation solutions that are tailored to user needs and preferences \cite{xu2019toward}.
One of the primary benefits of HCA is increased efficiency and productivity. When automation systems are designed with the user in mind, they can streamline workflows, reduce manual errors, and free up time for higher-value tasks \cite{davenport2016just}. For example, in scientific research, an HCA-designed automation solution for data analysis and visualization can help researchers quickly identify patterns and insights, accelerating the discovery process. In education, HCA-driven automation tools can personalize learning content and provide immediate feedback to students, enabling them to learn at their own pace and style.
Moreover, HCA can foster a culture of innovation and continuous improvement through the empowerment of users to actively participate in the automation process. When users feel that their needs and feedback are valued, they are more likely to embrace new technologies and contribute ideas for further optimization \cite{rahwan2019machine}. This collaborative approach can lead to the development of more effective and user-friendly automation solutions that address real-world challenges and drive business value. For instance, in organizations and NGOs, HCA can enable employees to identify opportunities for automation in their daily tasks and work with developers to create solutions that enhance their productivity and job satisfaction.

HCA can also contribute to increased job satisfaction and reduced stress levels among employees by providing automation tools that are intuitive and support their work, rather than replacing it. A study by Manyika et al. \cite{manyika2017future} found that employees in organizations that adopted a human-centered approach to automation reported higher levels of job satisfaction and engagement compared to those in organizations with a more technology-centric approach. Designing automation solutions that complement human skills and expertise, HCA can help create a more fulfilling and rewarding work environment.
Furthermore, HCA can play a crucial role in helping individuals and organizations stay competitive in the era of rapidly advancing AI and automation. Democratizing access to automation technologies through open-source solutions and user-friendly interfaces, HCA can lower the barriers to entry and enable a wider range of users to benefit from these technologies. This can help reduce the risk of job displacement and empower individuals to continuously upgrade their skills and adapt to new roles and responsibilities.
In summary, the benefits of HCA extend beyond productivity gains and cost savings. Prioritizing user needs, fostering collaboration, and promoting innovation, HCA can help create a more human-centric future of work, where automation technologies augment and support human capabilities rather than replacing them. As organizations across various domains increasingly adopt automation, embracing an HCA approach will be crucial to ensure that these technologies are developed and deployed in a way that benefits both individuals and society as a whole.

\section{Empirical Evidence Supporting Human-Centered Automation}

Several studies have investigated user preferences and pain points in interacting with automation systems, providing valuable insights for the development of HCA. 

\subsection{Studies highlighting user preferences and pain points in automation}

Several studies have studied the interaction with automation systems across various domains. One such study by Kaber and Endsley \cite{kaber2004effects} examined the effects of different levels of automation on user performance and situation awareness in a complex problem-solving task. The results showed that users preferred intermediate levels of automation that kept them involved in the decision-making process, rather than fully manual or fully automated control. This finding suggests that HCA should aim to strike a balance between human control and automated assistance, ensuring that users remain engaged and aware of the system's actions.
Another study by Parasuraman and Miller \cite{parasuraman2004trust} explored user trust and reliance on automation in a simulated flight task. The authors found that users' trust in the automation system was influenced by its reliability and transparency, with users preferring systems that provided clear explanations for their actions and allowed for human intervention when needed. This highlights the importance of designing automation systems that are transparent, explainable, and controllable, enabling users to build trust and effectively collaborate with the technology.

In the context of educational technology, a study by Roscoe et al. \cite{roscoe2015bridging} investigated learner preferences and experiences with an intelligent tutoring system that provided automated feedback and guidance. The results showed that learners appreciated the immediate and targeted feedback provided by the system but also valued the ability to ask questions and receive explanations from human instructors. This suggests that HCA in educational settings should aim to complement human instruction rather than replace it entirely and provide opportunities for learner-teacher interaction and collaboration.
In the healthcare domain, a study by Dogan et al. \cite{dogan2021ethics} explored nurses' experiences with an automated medication dispensing system. The authors found that while the system improved efficiency and reduced errors, nurses experienced frustration with its inflexibility and lack of consideration for their workflow and decision-making processes. This underscores the importance of involving end-users in the design and implementation of automation systems and ensuring that the technology adapts to the users' needs and preferences rather than the other way around.

These studies highlight the importance of considering user perspectives, needs, and preferences in the development of automation systems across various domains. Involving users in the design process, providing transparent and explainable systems, and allowing for human control and collaboration, HCA can create automation solutions that are more readily accepted, trusted, and effectively used by the intended users.

\subsection{Insights from human-computer interaction research applicable to automation}

Human-Computer Interaction (HCI) research offers a wealth of knowledge that can inform the design and development of HCA systems. One of the key principles of HCI is user-centered design, which emphasizes the importance of understanding user needs, goals, and contexts throughout the design process \cite{anderson1988user}. Applying user-centered design principles to automation, HCA can create solutions that are tailored to the specific requirements and preferences of the intended users, increasing the likelihood of adoption and satisfaction.
Another relevant insight from HCI research is the importance of feedback and transparency in interactive systems. Studies have shown that providing users with clear and timely feedback on the system's status, actions, and outcomes can enhance user understanding, trust, and control \cite{shneiderman2010designing}. In the context of automation, this suggests that HCA systems should provide users with meaningful feedback and explanations regarding the automated processes, enabling them to monitor and intervene as needed.

HCI research also highlights the value of participatory design, which involves actively engaging users in the design and development process \cite{muller1993participatory}. Involving users in the creation of automation solutions, HCA can ensure that the resulting systems are aligned with user needs and expectations and can benefit from the users' domain expertise and insights. This collaborative approach can lead to the development of more effective and user-friendly automation tools that address real-world challenges and drive innovation.
Furthermore, HCI research emphasizes the importance of adaptive and personalized interfaces that can accommodate the diverse needs and preferences of users. Studies have shown that interfaces that adapt to the user's skill level, context, and goals can improve user performance, satisfaction, and engagement \cite{hook1997cognitive}. In the context of HCA, this suggests that automation solutions should be designed to adapt to the user's needs and provide personalized support and guidance based on their individual characteristics and context.

Finally, HCI research also provides insights into the design of multimodal interfaces that can support more natural and intuitive human-machine interaction. Multimodal interfaces that combine different input and output modalities, such as speech, gestures, and visual displays, can enhance user experience and enable more efficient and effective communication between humans and machines \cite{oviatt2003multimodal}. Applying these principles to HCA, designers can create automation solutions that support more natural and seamless interaction, leveraging the capabilities of multimodal AI and LLMs to understand user behavior and intentions.
Leveraging the insights and principles from HCI research, HCA can create automation solutions that are more user-friendly, adaptive, and engaging, ultimately leading to better acceptance, trust, and effectiveness in supporting human activities across various domains.

\subsection{Case studies demonstrating the effectiveness of user-centric automation approaches}

Several case studies have demonstrated the effectiveness of user-centric automation approaches in various domains. One such example is the implementation of a human-centered RPA system at a global financial services company, as reported by Lacity and Willcocks \cite{lacity2016new}. The company involved users in the design and development of the RPA solution, conducting workshops and feedback sessions to ensure that the system met their needs and expectations. The resulting RPA system was well-received by users, who reported increased efficiency, reduced errors, and improved job satisfaction. This case study illustrates the benefits of adopting an HCA approach in the deployment of RPA technologies.
Another case study by Trippas et al. \cite{trippas2019learning} explored the application of a user-centric chatbot for customer service in the telecommunications industry. The chatbot was designed using a human-centered approach, incorporating user feedback and preferences throughout the development process. The results showed that the chatbot was effective in handling customer inquiries, reducing response times, and improving customer satisfaction. Moreover, the user-centric design of the chatbot led to increased trust and acceptance among users, who appreciated the system's transparency and ability to seamlessly transfer complex issues to human agents when needed.

In the healthcare domain, a case study by Gao et al. \cite{gao2021human} investigated the implementation of a human-centered AI system for medical diagnosis support. The system was designed in collaboration with healthcare professionals, taking into account their clinical workflows, information needs, and decision-making processes. The study found that the AI system improved diagnostic accuracy and efficiency while also enhancing the clinicians' trust and understanding of the system's recommendations. This case study highlights the importance of involving domain experts in the design of AI-based automation solutions to ensure their effectiveness and acceptance in real-world settings.
In the field of education, a case study by Kite et al. \cite{kite2020impact} explored the use of a user-centric intelligent tutoring system to support personalized learning in a high school mathematics course. The system was designed with the input of teachers and students, incorporating features such as adaptive feedback, gamification elements, and learner control. The results showed that students who used the intelligent tutoring system achieved higher learning gains and reported higher levels of engagement and motivation compared to those in a traditional classroom setting. This case study demonstrates the potential of user-centric automation approaches in enhancing educational outcomes and experiences.
These case studies provide empirical evidence of the effectiveness of HCA in creating automation solutions that are not only technically sound but also aligned with user needs and preferences. Prioritizing user involvement, transparency, and collaboration, HCA can help organizations realize the full potential of automation technologies while ensuring user satisfaction and acceptance across various domains.

\section{Towards a Framework for Human-Centered Automation}

Based on the empirical evidence and insights from HCI research, a set of key principles and guidelines for designing human-centric automation systems is proposed. 

\subsection{Key principles and guidelines for designing human-centric automation systems}

We propose a set of key principles and guidelines for designing human-centric automation systems. These principles aim to ensure that automation solutions are user-friendly, transparent, and adaptable, promoting effective collaboration between humans and machines across various domains.
Active user involvement in the design and development process is crucial. User research, workshops, and feedback sessions help to understand their needs, preferences, and contexts \cite{sanders2008co}. This user involvement ensures that the automation solution is tailored to the specific requirements of the intended users, increasing the likelihood of adoption and satisfaction.
Automation systems should be designed to provide clear and meaningful explanations of their actions, decisions, and outcomes. This transparency and explainability enable users to understand and trust the technology \cite{ribeiro2016should}. Transparency and explainability are critical for building user trust and acceptance, as they allow users to comprehend how the automation system works and why it makes certain recommendations or decisions.

Adjustable levels of automation should be implemented to allow users to choose the degree of automated assistance based on their preferences, skills, and the task at hand \cite{parasuraman2000model}. This adaptability ensures that users can maintain an appropriate level of control and engagement, preventing over-reliance on automation or feelings of disempowerment.
Automation solutions should be designed to integrate smoothly with users' existing workflows and tools to minimize disruption and cognitive load \cite{shneiderman2010designing}. This integration helps users to easily incorporate automation into their daily tasks, reducing the learning curve and increasing the likelihood of continued use.
User-friendly interfaces that are easy to navigate, understand, and control are imperative. Clear feedback and guidance should be provided to users \cite{nielsen1994usability}, reducing the cognitive effort required to interact with the system, making it more accessible to a wider range of users.

Effective collaboration between humans and machines should be fostered, leveraging the strengths of both to make informed decisions and drive innovation \cite{kamar2016directions}. This collaboration involves designing automation systems that support and augment human capabilities rather than replacing them entirely.
Regular collection of user feedback and monitoring of system performance is essential to identify areas for improvement and inform iterative design updates \cite{friedman2013value}. This ongoing refinement process ensures that the automation solution remains relevant, effective, and user-centric over time.
Applying these principles and guidelines, designers and developers can create automation solutions that prioritize user needs and preferences. This leads to increased adoption, satisfaction, and effectiveness across various domains. The incorporation of open-source components and standardized interfaces can further promote the accessibility and democratization of these technologies, enabling a wider range of users to benefit from the potential of HCA.

\subsection{Open challenges and future research directions}

While the proposed framework provides a foundation for designing human-centric automation systems, several open challenges and future research directions remain to be addressed. One key challenge is the need for more comprehensive and standardized methods for evaluating the usability and user experience of automation solutions \cite{dix2003human}. Existing evaluation methods may not fully capture the unique characteristics and impacts of automation technologies on users. Future research should focus on developing robust evaluation frameworks that consider the diverse needs of users and the specific contexts in which automation solutions are deployed. These frameworks should encompass both quantitative and qualitative measures, providing a holistic assessment of the user experience and the effectiveness of the automation solution. Collaboration between researchers from various disciplines, including HCI, psychology, and domain-specific fields, will be required to ensure that the evaluation methods are comprehensive, reliable, and valid.
Another important challenge is the ethical and social implications of automation, particularly in terms of job displacement, privacy concerns, and the potential for bias and discrimination. As automation technologies become more prevalent, it is crucial to consider the potential unintended consequences and develop strategies to mitigate them. Future research should explore ways to design automation systems that are not only user-centric but also socially responsible and ethically sound. This may involve developing guidelines and standards for the transparent and accountable use of automation technologies, as well as investigating the long-term impacts of automation on employment, skills development, and social inequality. Researchers should also work closely with policymakers, organizations, and the public to foster open dialogue and collaborative decision-making regarding the development and deployment of automation technologies.

Furthermore, the integration of automation solutions with existing tools, platforms, and infrastructures presents another significant challenge. Ensuring compatibility, interoperability, and scalability across different systems and domains is crucial for the widespread adoption and effectiveness of HCA. Future research should focus on developing standardized protocols, APIs, and architectures that enable seamless integration and data exchange between automation solutions and other technologies. Collaboration between academia, industry, and government stakeholders will be required to establish common standards and best practices for the design, development, and deployment of interoperable automation systems.
As automation technologies continue to advance and become more sophisticated, there is a need for ongoing research into the evolving roles of humans and machines in collaborative work environments. The increasing capabilities of AI systems may lead to new forms of human-machine interaction and collaboration, requiring a reevaluation of traditional task allocation and decision-making processes. Future studies should investigate how HCA can adapt to these changes and support the development of new skills, roles, and organizational structures that enable humans and machines to work together effectively. This may involve exploring new paradigms for task allocation, decision support, and performance monitoring, as well as developing training and education programs that prepare individuals for the future of work.

Finally, the long-term sustainability and adaptability of HCA systems is another important consideration for future research. As user needs, preferences, and contexts evolve, automation solutions must be designed to adapt and remain relevant over time. This may involve developing self-learning and self-optimizing mechanisms that can automatically adjust the system's behavior based on user feedback and changing requirements. Researchers should also investigate the environmental impact of automation technologies and develop strategies for designing and deploying sustainable and energy-efficient solutions.
Addressing these challenges and pursuing these research directions will require a collaborative and interdisciplinary effort from the research community, industry practitioners, policymakers, and the public. Working together and leveraging the insights and principles of HCA, we can create a future in which automation technologies are designed to empower and support humans across various domains, from scientific research and education to organizations, NGOs, and governments. This will not only lead to more effective and user-friendly automation solutions but also contribute to the development of a more inclusive, sustainable, and human-centric society.

\section{Conclusion}

In conclusion, the position paper argues for the importance of developing and adopting a Human-Centered Automation (HCA) approach to designing and implementing automation systems across various domains. The current state of automation, characterized by the limitations of rule-based RPA systems and the challenges of integrating Generative AI technologies, highlights the need for a more user-centric approach. The proposed framework for HCA provides a set of guiding principles and recommendations for designing automation systems that prioritize user needs, foster effective human-machine collaboration, and promote the accessibility and democratization of these technologies through open-source solutions.

The empirical evidence and insights from HCI research support the effectiveness of user-centric automation approaches in various domains, demonstrating the benefits of increased efficiency, productivity, innovation, job satisfaction, and reduced stress levels among users. Active user engagement in the design process, providing transparent and explainable systems, and enabling seamless integration with existing workflows, HCA can help overcome the challenges of non-intuitive interfaces and create automation solutions that are more readily adopted and trusted by users.
However, realizing the full potential of HCA requires addressing several open challenges and future research directions. These include the need for more comprehensive and standardized evaluation methods, the ethical and social implications of automation, the integration of automation solutions with existing technologies, the evolving roles of humans and machines in collaborative work environments, and the long-term sustainability and adaptability of HCA systems. A collaborative and interdisciplinary effort from researchers, practitioners, policymakers, and the public will be necessary to tackle these challenges.

The position paper calls for researchers and practitioners to embrace the HCA approach and actively contribute to its development and implementation. Prioritizing user needs, providing intuitive interfaces, and leveraging the capabilities of AI, HCA can help create a more accessible, user-friendly, and empowering future of automation. This approach not only benefits individuals and organizations but also contributes to the development of a more inclusive and sustainable society, where automation technologies augment and support human capabilities rather than replace them.
Ultimately, the success of HCA relies on the collective effort and commitment of all stakeholders involved in the development and deployment of automation technologies. Working together and embracing the principles of HCA, we can shape a future in which automation empowers people across various domains, fosters innovation and creativity, and helps individuals and organizations remain competitive in the era of rapidly advancing AI. It is our responsibility to ensure that the benefits of automation are distributed equitably and that the technology is developed and used in a way that promotes human values, well-being, and social progress.

\subsection{Summary of key points and recommendations}

The position paper presents a compelling case for the adoption of Human-Centered Automation (HCA) and underscores the need for a more user-centric approach. The proposed framework for HCA outlines key principles and guidelines for creating automation solutions that prioritize user needs, foster effective human-machine collaboration, and promote accessibility through open-source initiatives.
The empirical evidence and insights from HCI research highlight the benefits of user-centric automation approaches, including increased efficiency, productivity, innovation, and job satisfaction. Active user involvement in the design process, ensuring transparency and explainability, and seamlessly integrating with existing workflows, HCA can help overcome the challenges of non-intuitive interfaces and create automation solutions that are more readily adopted and trusted by users.
However, the paper also acknowledges the open challenges and future research directions that need to be addressed to fully realize the potential of HCA. These include developing more comprehensive and standardized evaluation methods, addressing the ethical and social implications of automation, seamlessly integrating automation solutions with existing technologies, adapting to the evolving roles of humans and machines in collaborative work environments, and ensuring the long-term sustainability and adaptability of HCA systems.
The paper recommends that researchers and practitioners actively embrace the HCA approach and contribute to its development and implementation. Prioritizing user needs, providing intuitive interfaces, and leveraging the capabilities of advanced AI technologies, HCA can help create a more accessible, user-friendly, and empowering future of automation that benefits individuals and organizations across various domains.

\subsection{Call to action for researchers and practitioners}

In light of the compelling case for Human-Centered Automation presented in this position paper, we call upon researchers and practitioners from various disciplines, including HCI, AI, RPA, and domain-specific fields, to actively embrace and contribute to the development and implementation of HCA principles and guidelines. Working together and leveraging our collective expertise, we can create automation solutions that truly prioritize user needs, foster effective human-machine collaboration, and promote the accessibility and democratization of these technologies.

We urge researchers to focus on addressing the open challenges and future research directions outlined in this paper, such as developing more comprehensive and standardized evaluation methods, investigating the ethical and social implications of automation, and exploring new paradigms for human-machine interaction and collaboration. This will require a collaborative and interdisciplinary effort, bringing together experts from academia, industry, and government to share knowledge, best practices, and innovative ideas.

Practitioners, including designers, developers, and domain experts, play a crucial role in translating HCA principles into practice. We encourage practitioners to actively engage with users throughout the design and development process, seeking their input, feedback, and validation at every stage. Involving users as co-creators and collaborators, we can ensure that the resulting automation solutions are not only technically sound but also aligned with users' needs, preferences, and contexts.
Furthermore, we call upon organizations, policymakers, and funding agencies to support and invest in HCA research and development initiatives. This may involve providing resources for user-centered design activities, promoting open-source collaboration, and establishing policies and guidelines that prioritize user needs and ethical considerations in the development and deployment of automation technologies.

Ultimately, the success of Human-Centered Automation relies on the collective effort and commitment of all stakeholders involved in shaping the future of automation. Embracing the HCA approach and working together towards a common goal, we can create a future in which automation technologies empower and support humans across various domains, fostering innovation, creativity, and social progress. It is our responsibility to ensure that the benefits of automation are distributed equitably and that the technology is developed and used in a way that promotes human values, well-being, and flourishing.
Let us seize this opportunity to make a positive impact and pave the way for a more human-centric, accessible, and empowering future of automation. Together, we can build a world in which humans and machines work side by side, leveraging each other's strengths and compensating for each other's limitations, to achieve greater efficiency, innovation, and well-being for all.

\bibliographystyle{ACM-Reference-Format}
\bibliography{new}

\end{document}